\documentclass[rsi,amsmath,amssymb,reprint,floatfix]{revtex4-1}
\usepackage[utf8]{inputenc}
\usepackage{graphicx}
\usepackage{dcolumn}
\usepackage{bm}
\usepackage{float}
\usepackage{footmisc}
\usepackage[utf8]{inputenc}
\usepackage{xcolor}
\usepackage[T1]{fontenc}
\usepackage{mathptmx}
\usepackage{xspace}

\usepackage[
disable
]{endfloat}
\begin{document}

\preprint{AIP/123-QED}

\title[Tuneable homogeneous kG magnetic field production using permanent magnets]{Tuneable homogeneous kG magnetic field production using permanent magnets}

\author{Danielle Pizzey}
\email{danielle.boddy@durham.ac.uk}
\affiliation{Joint Quantum Centre (JQC) Durham-Newcastle, Department of Physics, Durham University, South Road, Durham, DH1 3LE, United Kingdom}

\date{\today}

\begin{abstract}

We present a permanent magnet arrangement that can achieve a tuneable axial magnetic field from 1.80(8)~--~2.67(2)~kG along the $z$-axis, where ($x$,$y$) = (0,0), with a rms field variation of less than 1\% over a length of 25~mm. The instrument consists of an arrangement of off-the-shelf N42 neodymium-iron-boron (NdFeB) axially magnetized ring magnets, of varying outer and inner diameters. The magnets are organized into four cylindrical brass holders, whose relative separation can be manipulated to achieve the desired magnetic field strength and homogeneity over the region of interest. We present magnetic field computations and  Marquardt-Levenberg fits to experimental data and demonstrate excellent agreement between theory and experiment. The apparatus has been designed to accommodate a cylindrical atomic vapor cell of length 25~mm and diameter 25~mm to lie within the bore of the ring magnets. The design has a clear aperture of 20~mm in the $x$-$y$ plane along the $z$-axis, which opens up new avenues for imaging exploration in atomic spectroscopy.

\end{abstract}

\maketitle

\section{\label{sec:intro}Introduction}

There are a vast number of situations in experimental atomic and optical physics where magnetic fields are utilized, from a Zeeman slower necessary to slow and cool an atomic beam \cite{doi:10.1063/1.4945795, doi:10.1063/1.3600897}, to magneto-optical filters (MOFs) that use thermal atomic vapors in magnetic fields \cite{Zentile:15}. MOFs find a wide range of applications in other disciplines including e.g. quantum key distribution \cite{doi:10.1063/1.2387867}; optical isolators \cite{Weller:12}; atmospheric LIDAR \cite{Yang2011, Xia:s}; and laser frequency stabilization \cite{ISI:000496295700001,Miao:11}. In certain cases there is the need to produce large, uniform magnetic fields (i.e. > 1 kG) in order to study a specific branch of physics, in particular thermal vapor experiments working in the hyperfine Paschen-Back regime \cite{doi:10.1088/1361-6455, doi:10.1080/09500340.2017.1377308}. It is not an easy feat to produce large uniform magnetic fields: a work-around is typically to reduce the physical dimensions of the interrogation region over which the experiments are performed \cite{PhysRevA.84.063410,Sargsyan:14}. However, this is not applicable to all situations and the need to produce large magnetic fields over increasingly large length-scales is ever pressing. A particular example is in solar flare forecasting \cite{10.1007/BF02515781}: the Sun is imaged through an optical system referred to as a solar filter \cite{Cimino1968,Cacciani:75}. The solar images taken are used to extract the line-of-sight solar magnetic field that is necessary for the forecasting model \cite{Korsos2018, Korsos2020a,Erdelyi21}. The optimum parameters for a solar filter of a particular atomic species depends on numerous factors, including the axial magnetic field, vapor cell length and vapor temperature. The optimum parameters can be determined theoretically using a modified version of ElecSus, which is a software package that calculates the electric susceptibility of an atomic medium \cite{Zentile2015b}, and it is these parameters that set the physical constraints on the magnetic field strength, geometry and dimensions of the solar filter. Often, solar filters require axial magnetic fields of several kilogauss that are uniform over length-scales of tens of millimeters along the $z$-axis \cite{Stangalini2018}. Additionally, since these filters are used in imaging, large clear apertures are required in the $x$-$y$ plane, which adds to the complexity of designing a system that can produce large magnetic fields. It is these constraints that we have complied with in this work.

Magnetic fields can be produced via an electromagnet, which uses a coil of wire and a current supply, or by a permanent magnet. There are many pros and cons to choosing either method. One particular benefit to using the electromagnet is the ability to tune the magnetic field by simply changing the current supplied to the coil; this is advantageous when testing a new idea or characterizing an experiment. However, to produce magnetic fields of several kG using electromagnets, cooling is typically required to prevent over-heating of the coils if used continuously \cite{Ilja2014,doi:10.1063/1.2785157, Gerhardt:18}. On the flip side, permanent magnet arrangements do not have any power requirements, hence no cooling, nor do they suffer from magnetic field noise due to current fluctuations originating from the power supplies. A ``set and forget'' approach is ideal when making a system transportable and controlled remotely, as this reduces the number of potential technical failures during field-work. It is for this reason that we have opted to pursue the permanent magnet route. With appropriately designed hardware, we address the tuneability problem and demonstrate field homogeneity over the required length-scale. 

The remainder of the paper is organized as follows: in Section~\ref{sec:mag_field_comp} we present the magnetic field computation; Section~\ref{sec:hardware_criteria} discusses the functionality of the hardware and limitations to the maximum achievable magnetic field; we illustrate the performance of the assembly in Section~\ref{sec:results}; and finally conclusions are drawn and an outlook provided in Section~\ref{sec:conclusion}.

\section{\label{sec:design_overview}Design Overview}

The magnet geometries described in this paper were designed specifically for a potassium solar filter. The requirements are: a maximum axial magnetic field of $\approx$~2.7~kG; a field homogeneity over 25~mm along the $z$-axis; a clear imaging window of a minimum of 20~mm diameter in the $x$-$y$ plane; minimal contribution from the $x$- and $y$-components of the magnetic field across the imaging window; and $\approx$~1~kG tuning range of the magnetic field strength along the $z$-axis.

\subsection{\label{sec:mag_field_comp}Magnetic field computation}

Finding a permanent magnet arrangement that abides by the constraints stated in section~\ref{sec:design_overview} is a complex task. There are many software packages available that enable the magnetic field to be determined, however we chose to use Magpylib~\cite{ORTNER2020100466}, which is a python package for magnetic field computation. The Magpylib package is open source and provides visual aids of the magnet geometry, which assisted the design of hardware mount. 

The criteria of the instrument, outlined in section~\ref{sec:design_overview}, state there must be a clear aperture of 20~mm in diameter in the $x$-$y$ plane along the magnetic field axis (i.e. $z$-axis) to permit solar imaging through the atomic vapor cell. Since the magnetic field is a vector quantity, $\bf{B}\,\equiv\,(B_{x},B_{y},B_{z})$, using magnets with cylindrical symmetry will make the design simpler to facilitate. We focus our studies on ring magnets with opposite polarity on opposite faces as this produces an axial magnetic field. 

\begin{figure}[tbh!]
\begin{center}
\includegraphics[width=85mm,clip=true,trim = 2mm 0mm 0mm 0mm]{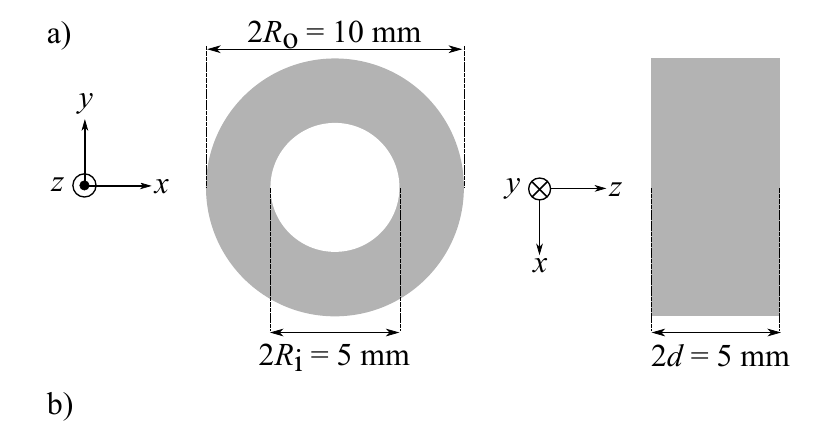}
\includegraphics[width=85mm,clip=true,trim = 10mm 0mm 0mm 0mm]{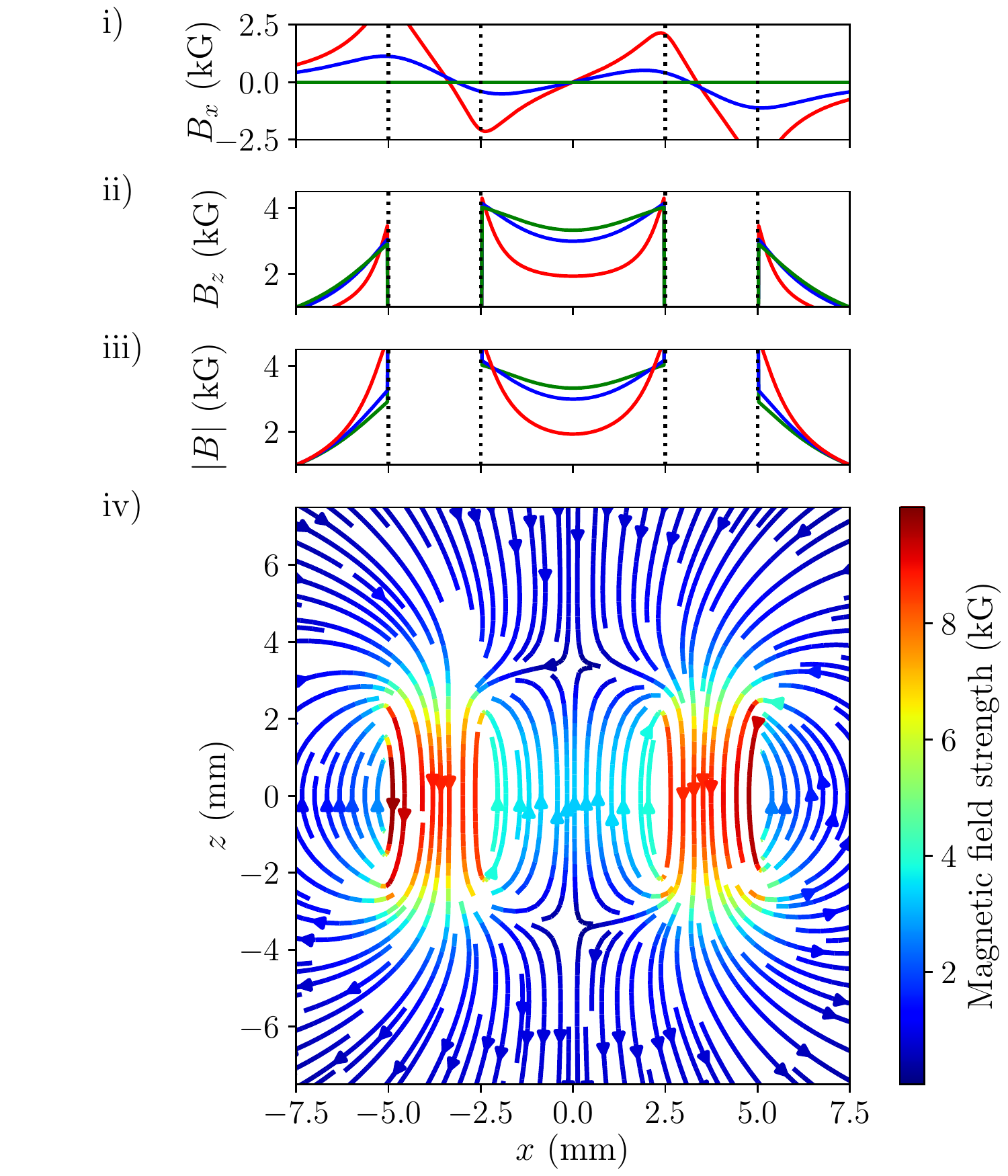}
\end{center}
\vspace{0mm}
\caption{Simple picture of a) a ring magnet with dimensions 2$R_\mathrm{o}$~=~10~mm, 2$R_\mathrm{i}$~=~5~mm and 2$d$~=~5~mm and remanence field $B_{0}$~=~12.8 kG, typical of N42 grade NdFeB magnets and the associated b) magnetic field profile calculated using Magpylib \cite{ORTNER2020100466}. Slices of the magnetic field components i) $B_{x}$ and ii) $B_{z}$ as a function of $x$ taken at $z$~=~0~mm (green), $z$~=~1~mm (blue), and $z$~=~2~mm (red). Vertical black dotted lines represent the physical profile of the magnet, i.e. $\pm$~5~mm illustrates the outside diameter of the ring, whereas $\pm$~2.5~mm illustrates the inside diameter of the ring. Also shown is the total magnitude of the magnetic field given by Eq.~\ref{eq:field_mag} as a function of $x$ taken at $z$~=~0~mm (green), $z$~=~1~mm (blue), and $z$~=~2~mm (red).  iv) Magnetic field profile displayed in the $x$-$z$ plane illustrated in magnitude by color and direction by arrows.}
\label{fig:radial}
\end{figure}

In order to design an appropriate magnet assembly we must first understand the magnetic field profile of a single ring magnet. 
Figure \ref{fig:radial} illustrates the magnetic field profile of a ring magnet with outer diameter 2$R_\mathrm{o}$~=~5~mm, inner diameter 2$R_\mathrm{i}$~=~2.5~mm, length 2$d$~=~5~mm, as shown in Fig.~\ref{fig:radial}\,a), and remanence field $B_{0}$~=~12.8~kG, which is typical of N42 grade NdFeB magnets. Although this ring magnet does not meet our constraints, we can learn a lot from the figure. In Fig.~\ref{fig:radial}\,b)\,iv), the magnetic field profile is shown in the $x$-$z$ plane and, due to cylindrical symmetry, we can assume the profile is identical in the $y$-$z$ plane also. The arrows demonstrate the field direction, while the color of the arrow represents the magnitude of the field. Inside the physical profile of the magnet, the field is strongest. Within the bore of the ring magnet we see that the field is the opposite to that inside the magnet and a lot weaker, as evidenced by the change in color of the arrow. Additionally, the field lines within the bore of the ring magnet are not evenly spaced along the $x$-axis and display curvature, indicating that the field along $x$ is not homogeneous. This is evident in Fig.~\ref{fig:radial}\,b)\,i) and ii), which show the $B_{x}$ and $B_{z}$ component of the magnetic field as a function of $x$ for three $z$ positions. At $x$~=~0~mm, we see that $B_{x}$~=~0 for all $z$ positions. However, as we translate along $x$, the argument that $B_{x}$~=~0 is no longer valid except for the case when $z$~=~0~mm. The further we move from ($x$,$z$) = (0,0), the larger the $x$ component of the field becomes. Additionally, we see that $B_{z}$ begins to reduce as we move further from $z$~=~0~mm. In practice it is difficult to measure an individual component of the magnetic field and it is typically the magnitude of the total field that is measured at a particular point in space. Figure~\ref{fig:radial}~b)~iii) illustrates the magnitude of the total field as a function of $x$ for three $z$ positions discussed in Fig.~\ref{fig:radial}~b)~i)~and~ii). The magnitude of the total field along $x$ is not homogeneous and displays curvature similar to that of the $B_{z}$ component of the magnetic field, since this component is dominant over $B_{x}$ and $B_{y}$ in this design. The details of why this occurs will not be discussed in detail here, as it is beyond the scope of this paper, however a thorough description and investigation can be found in ref~\cite{Trenec:11}.

This simple picture aids our understanding of the magnetic field profile of ring magnets. There are three take home messages that we learn from Fig.~\ref{fig:radial}. First, there is always a radial component of the magnetic field when ($x$,$y$) $\ne$ (0,0). Nevertheless, with an appropriately designed magnet assembly, the radial component can be made to be significantly smaller than the axial field component, so much so that we can neglect it in our computations. It is for this reason that we can assume that a cylindrical magnet with uniform magnetization, whose axis is aligned with the $z$~=~0 axis and $B_{x}$~=~$B_{y}$~=~0, has the following $z$-component of the magnetic field \cite{LeeWellerthesis},

\begin{equation}
    B_{z}(z) = \frac{B_{0}}{2}\left( \frac{z-z_{0}+d}{\sqrt{(z-z_{0}+d)^{2}+R^{2}}}- \frac{z-z_{0}-d}{\sqrt{(z-z_{0}-d)^{2}+R^{2}}} \right).
    \label{eq:field}
\end{equation}
Here, 2$d$ is the length of the cylinder, $R$ is its radius, $B_{0}$ is the remanence field, and $z_{0}$ is the position of the cylinder's centre along the $z$-axis. The magnetic field for a ring is calculated using the principle of superposition. For a ring magnet with outer radius $R_\mathrm{o}$ and inner radius $R_\mathrm{i}$ the field is that from a cylinder of radius $R_\mathrm{o}$ minus the field from a cylinder with $R_\mathrm{i}$. 

Second, to produce field homogeneity along $x$, and similarly $y$, the physical dimensions of the magnets used should significantly exceed the area of required field homogeneity. For example, if we require a clear aperture of diameter 4~mm in the $x$-$y$ plane with field homogeneity along $z$ and minimal radial components, the magnet ring specifications illustrated in Fig.~\ref{fig:radial}\,a), particularly the dimension 2$R_\mathrm{i}$~=~5~mm, would be inadequate as there are huge field variations when $x$~>~+1~mm and $x$~<~--1~mm. This ring magnet is more likely suited for a region of interest of diameter 2~mm in the $x$-$y$ plane.

Third, there will always be a reduction of the field strength and a flip of the field direction on the axis of the ring magnet upon exit of the bore of the magnet. This occurs at approximately $\pm$~3~mm in Fig.~\ref{fig:radial}\,b)\,iv). Thus it is pertinent to have the length of the ring magnet exceeding the desired homogeneity length along the $z$-axis.

Using Magpylib, we were able to address the three take home messages specific to our case. We prioritized our search on ring magnets with inner diameter greater than 25~mm for two reasons: we needed to fit a 25~mm diameter vapor cell inside the central bore of the magnets and; for our imaging purposes, we required a clear aperture of diameter 20~mm with minimal magnetic field variations across $x$-$y$ and along $z$. Additionally, since we require field homogeneity along the $z$-axis of 25~mm (the length of our vapor cell), the length (or thickness) of the ring magnet must exceed 25~mm. In an ideal scenario, the length of the ring magnet would be infinite, but this is not feasible when making an experimental instrument. An alternative to using excess magnet length is to increase the magnetic field at the boundaries of the region of interest, in our case at the edges of the vapor cell (i.e. at $z$~=~$\pm$~12.5~mm). The field can be increased by either using an additional ring magnet on the boundaries that has a higher remanence field (i.e. higher grade/strength) or by using a greater quantity of magnetic material. We chose the latter route as working with higher strength magnets are more difficult to manoeuvre and more dangerous to handle.

\begin{figure}[tbh!]
\begin{center}
\includegraphics[width=85mm,clip=true,trim = 0mm 0mm 0mm 0mm]{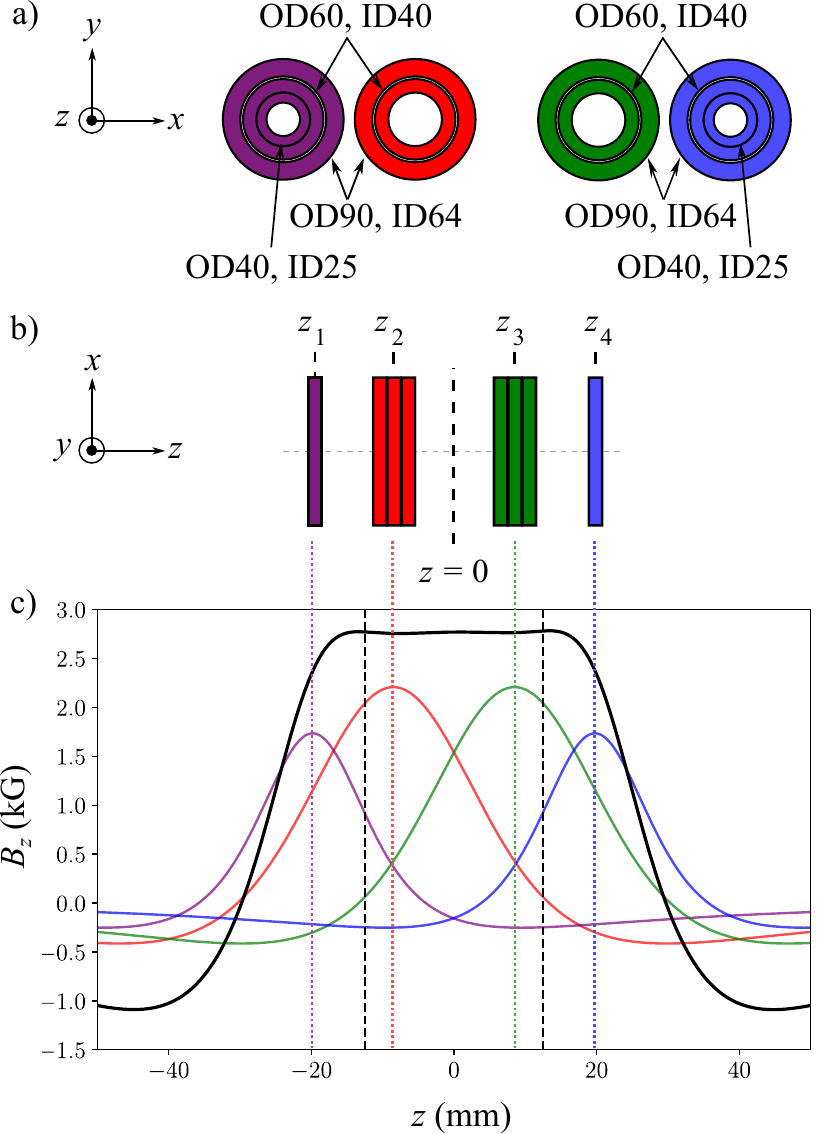}
\end{center}
\vspace{0mm}
\caption{a) Illustration of the concentric magnet arrangement in the $x$-$y$ plane used in the computation model. "OD" and "ID" correspond to the outer and inner diameters of the magnets, respectively, with the number succeeding either OD or ID indicating the dimension in millimeters. The design can be thought of as four magnet groups, illustrated by use of four colors: purple; red; green; and blue. The red and green magnet groups consist of three magnet layers, with each layer containing two concentric magnets. The purple and blue magnet groups consist of one magnet layer with three concentric magnets. b) Illustration of the magnet arrange in the $x$-$z$ plane. The magnet groups are positioned accordingly: purple positioned at $z_1$; red at $z_2$; green at $z_3$; and blue at $z_4$. c) Theoretical calculation of the axial magnetic field, $B_{z}$, using Eq.~\ref{eq:field}. The field from each individual magnet group is shown and illustrated via their corresponding color, with the magnet group position shown using dotted lines and $z_{i}$ label, where $i$~=~1,~2,~3, or 4. The total magnetic field is a superposition of the fields from the four magnet groups and is shown via a solid black line. Also shown are dashed black vertical lines that define a boundary  between which is the region of interest for the field homogeneity, where for this case $B_{\mathrm{mean}}$ = 2.8~kG with a root-mean-square field variation of less than 1\%.
\label{fig:Magpy_output}}
\end{figure}

Figure \ref{fig:Magpy_output} illustrates a ring magnet geometry that meets the criteria outlined in section~\ref{sec:design_overview}. The magnets used in this computation are commercial off-the-shelf neodymium-iron-boron (NdFeB) (N42) ring magnets of thickness 5~mm. We use three sizes: outer diameter (OD) 90~mm, inner diameter (ID) 64~mm;  OD60, ID40; OD40, ID25. The arrangement and positioning of the magnets is shown in Fig.~\ref{fig:Magpy_output}~a) and b). The design is composed of two halves that have exact-reflective symmetry about the $z$ = 0~mm mirror line: one half contains the magnets positioned at $z_1$ and $z_2$ (colored purple and red, respectively); while the other half contains the magnets positioned at $z_3$ and $z_4$ (colored green and blue, respectively). The red and green magnet groups comprise of three magnet layers, with each layer consisting of two concentric ring magnets (OD90, ID64 and OD60, ID40) with the same polarity. The polarity between each adjacent face of the layers is opposite, hence the layers attract, resulting in zero separation between the layers. The length, or thickness, of both the red and green group is 15~mm; this exceeds the edges of the vapor cell that are positioned at $z$~$\pm$~12.5~mm. Furthermore, the smallest inner diameter magnet used in these groups is 40~mm, which can also accommodate a vapor cell of diameter 25~mm.

The purple and blue magnet groups comprise of one magnet layer composed of three concentric ring magnets (OD90, ID64; OD60, ID40; and OD40, ID25) with the same polarity: these magnet groups are used to boost the field at the vapor cell boundaries, as discussed in the third take home message. In total there are 18 magnets in the system \footnote{Note: Should the reader wish to reproduce this design, custom made magnets can be used instead to minimize magnet layering.}. The axial magnetic field for this arrangement is calculated using Eq.~\ref{eq:field} and the total field is shown in Fig.~\ref{fig:Magpy_output}~c), as well as the individual contributions from the four magnet groups. The black dashed vertical lines define the positions of cylindrical vapor cell faces. The mean axial magnetic field between the two cell faces is $B_{\mathrm{mean}}$~=~2.8~kG with a root-mean-square field variation of less than 1\%. By changing the positions ($z_1$, $z_2$, $z_3$ and $z_4$) with respect to $z$~=~0, we can change the magnetic field strength while still maintaining field homogeneity over the region of interest. As a rule of thumb, the red and green magnet groups control the strength of the field, while the purple and blue magnet groups tend to control the extent of the homogeneity. 

\begin{figure}[tbh!]
\begin{center}
\includegraphics[width=85mm,clip=true,trim = 0mm 0mm 0mm 0mm]{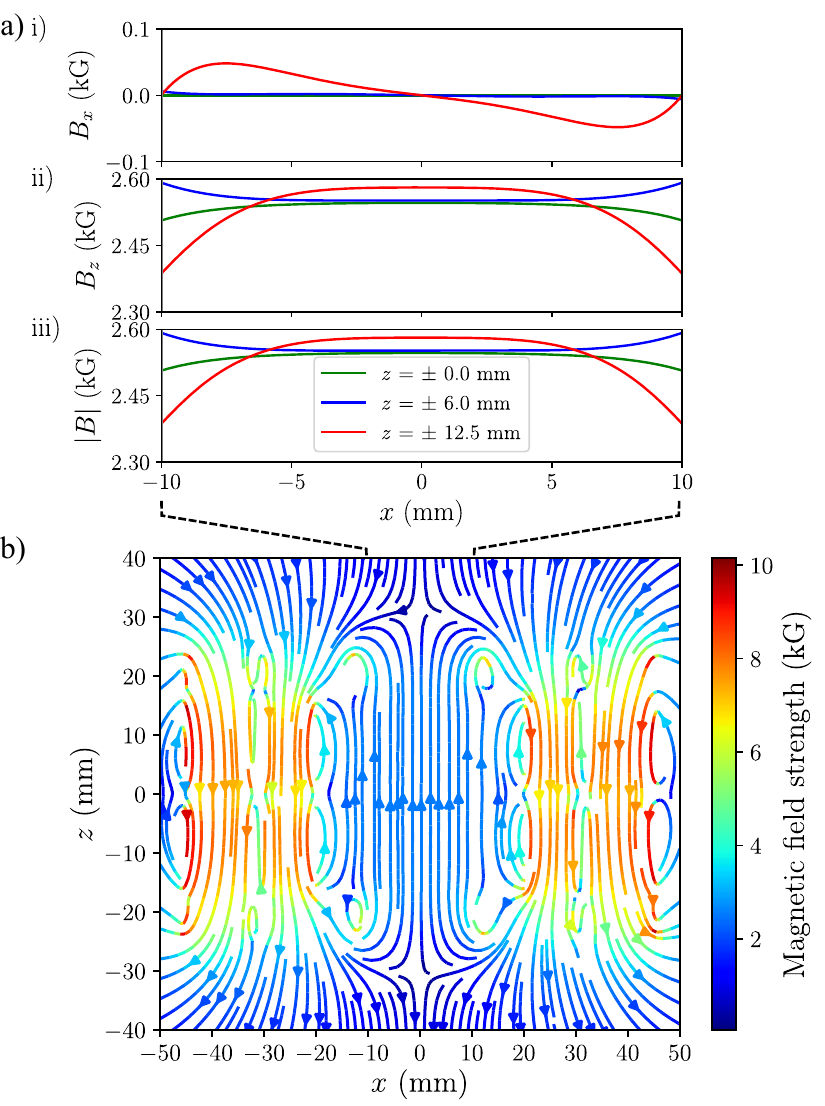}
\end{center}
\vspace{0mm}
\caption{Magpylib magnetic field computation for the magnet assembly shown in Fig.~\ref{fig:radial}. a) Slices of the magnetic field components i) $B_{x}$ and ii) $B_{z}$ as a function of $x$ taken at $z$~=~0.0 mm (green),  $z$~=~6.0~mm (blue) and $z$~=~12.5~mm (red). The vertical, black dashed lines illustrate the clear imaging window we require. The $B_{z}$ component greatly exceeds that of the $B_{x}$, and correspondingly $B_{y}$, components of the magnetic field. From Eq.~\ref{eq:field_mag}, the contribution is less than 1\%. Bottom: Magnetic field displayed in the $x$-$z$ plane illustrated in magnitude by color and direction by arrows.  
\label{fig:WS_Magpy}}
\end{figure}

Figure~\ref{fig:WS_Magpy} illustrates the magnetic field profile in the $x$-$z$ plane determined using Magpylib for the magnet arrangement shown in Fig.~\ref{fig:Magpy_output}. As shown in Fig~\ref{fig:WS_Magpy}\,b), the magnetic field strength is at its strongest inside the magnets, as expected. Within the bore of the magnets, where the vapor cell resides, the field lines are uniform across $x$, which is evident in Fig.~\ref{fig:WS_Magpy}\,a)\,i) and ii). Shown are the $B_{x}$ and $B_{z}$ components as a function of $x$ for three $z$ positions, which encompass the region of interest. Since the magnitude of the total magnetic field, $B_\mathrm{{total}}$, is given by

\begin{equation}
   | B_{\rm{total}}| = \sqrt{(B_{x})^{2} + (B_{y})^{2} + (B_{z})^{2}},
    \label{eq:field_mag}
\end{equation}
the contribution of $B_{x}$ and $B_{y}$ is less than 1\%. This is evident in Fig~\ref{fig:WS_Magpy}\,a)\,iii) where the magnitude of the field along $x$ echoes the $B_{z}$ profile shown in Fig~\ref{fig:WS_Magpy}\,a)\,ii).

\subsection{\label{sec:hardware_criteria}Hardware criteria}

\begin{figure}[tbh!]
\begin{center}
\includegraphics[width=8.5cm,clip=true,trim = 0mm 0mm 0mm 0mm]{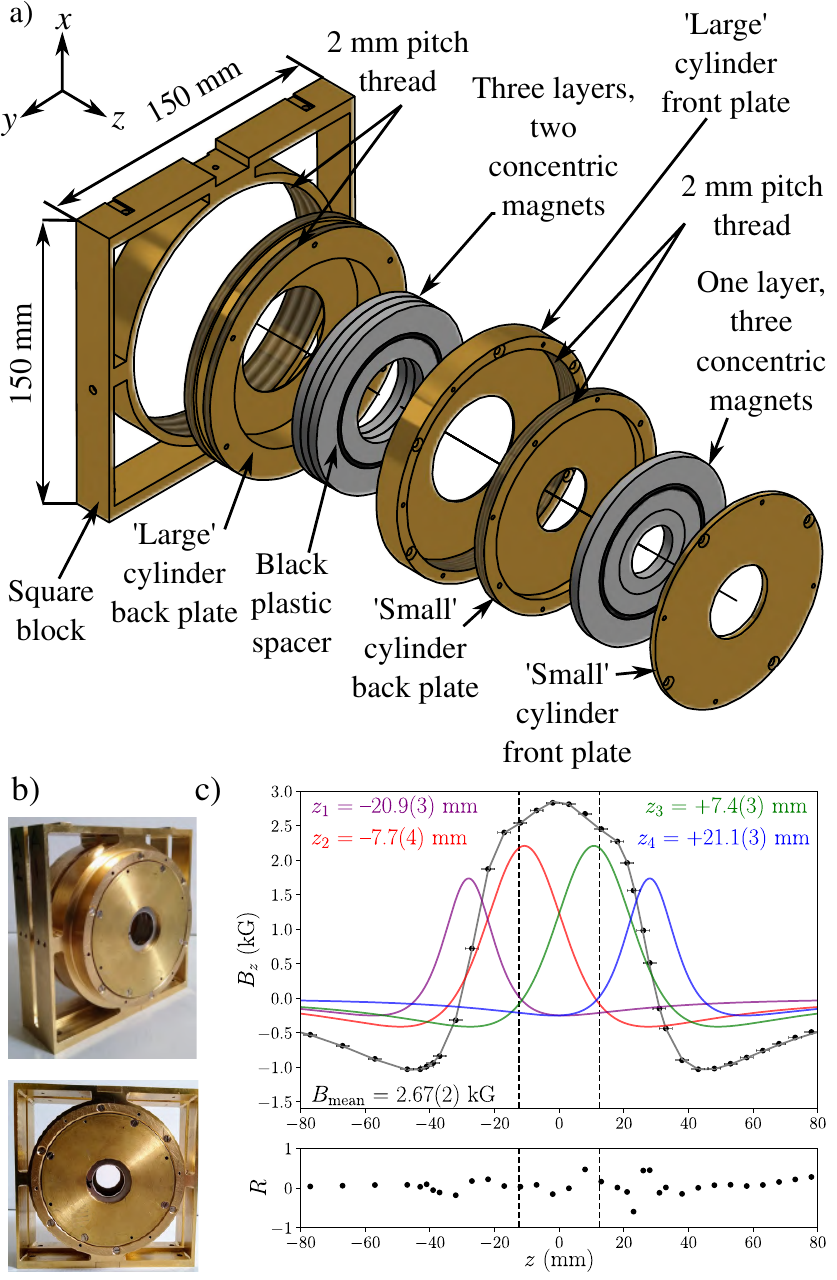}
\end{center}
\caption{a) An exploded view schematic illustration of one half of the magnet holder, which is composed of a square block, a ``large'' cylinder and a ``small'' cylinder. Also shown are the magnets and black plastic spacers. b) Photographs of the complete assembly. c) Measured magnetic field profile $B_{z}$ (black dots) when the assembly is set to give maximum field. The average magnetic field over the region of interest, illustrated via the black dashed lines, is $B_{\mathrm mean}$~=~2.67~kG with a root-mean-square deviation of 20 G over this region, i.e. 1\% variation. Also shown are the individual contributions from the four magnet groups that are colored accordingly, as outlined in figure~\ref{fig:Magpy_output}, with the positions and corresponding errors extracted from the Marquardt-Levenberg fit (grey solid line) \cite{Hughes2010} using Eq. \ref{eq:field} and the experimental data. The normalized residuals, $R$, are also shown.
\label{fig:magnet_geometry}}
\end{figure}

The magnet holder consists of two halves, that are mirror images of one another, and can be separated to permit insertion of a vapor cell to lie within the bore of the magnets. Each half consists of a square block and two cylinders, as shown in Fig.~\ref{fig:magnet_geometry}~a): one we define as the ``large" cylinder; the other we define as the ``small" cylinder. The large cylinder contains either the green or red magnet group, as shown in Fig.~\ref{fig:Magpy_output}. The small cylinder contains either the purple or blue magnet group.

The large cylinder back plate is threaded on the outside with a pitch of 2~mm and screws into the square block, permitting axial translation of 2~mm per rotation. The thickness of the square block sets how much the large cylinder can translate along the $z$-axis -- a square block thickness of 20~mm is sufficient to produce the fields investigated in this paper. The front plate of the large cylinder is threaded on the inside of its bore with a pitch of 2~mm. The back plate of the small cylinder is threaded on the outside which enables the small cylinder to be screwed into the internal bore of the large cylinder front plate. This permits an axial translation of the small cylinder of 2~mm per one full 2$\pi$ rotation, with respect to the large cylinder. This pitch is sufficient to provide the tuning control we require. Photographs of the complete assembly are shown in Fig.~\ref{fig:magnet_geometry}~b). In the bottom photograph the large imaging window is clearly visible.

Figure~\ref{fig:magnet_geometry}~c) illustrates the measured maximum field strength and the field profile along the $z$-axis, for the magnets and hardware used in this design \footnote{Note: The thickness of the front and back plates should be taken into account when simulating the magnetic field in Magpy (see Supplementary Materials for a further discussion).}. The mean magnetic field along the $z$-axis over the region of interest of $z~=~\pm$12.5~mm is $B_\mathrm{{mean}}$~=~2.67~kG with a root-mean-square field variation of 20~G, which is on the order of $\approx$ 1\% variation. Also stated are the positions of the magnet groups in their respective colors. 
Below, normalized residuals are also displayed.

The material chosen for the magnet holder is brass. Brass is non-magnetic and is durable enough to withstand cross-threading while screwing the cylinders closer together or further apart when tuning the magnetic field strength. To vary the cylinders positions, additional holes were included on each cylinder front plates such that custom tools could be inserted and used to rotate the cylinders and, hence, translate them. CAD drawings for these tools are included in the Supplementary Materials.

The assembly has been designed such that N42 grade NdFeB magnets can be used and achieve the fields necessary for our investigation. Magnets of this strength are easier to control and position, compared to N52 grade magnets, however, assembling the magnets into the cylinders must be approached with caution. An assembly guide has been provided in the Supplementary Materials, including CAD drawings for assembly jigs and photographs for visual aids.

\section{\label{sec:results}Results}

In the remainder of the paper, we demonstrate how we can manipulate the magnetic field profile by varying the positions of the magnet groups using the hardware discussed in section~\ref{sec:hardware_criteria}. We compare our measurements to theoretical calculations using Eq.~\ref{eq:field}.

\begin{figure}[tbh!]
\begin{center}
\includegraphics[width=85mm,clip=true,trim = 0mm 0mm 0mm 0mm]{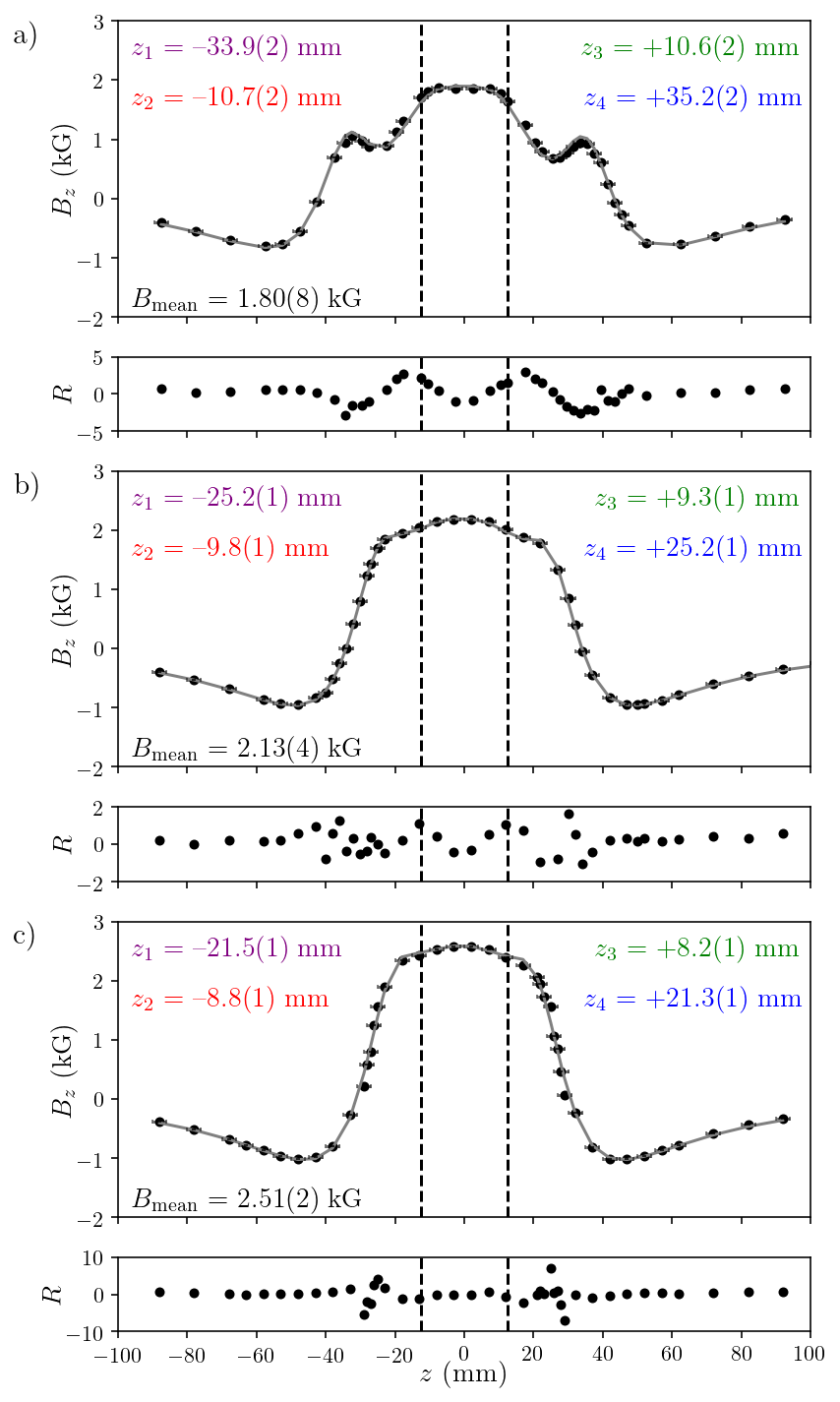}
\vspace{0cm}
\end{center}
\caption{Measured axial magnetic field (black dots) with an optimized fit (solid grey line) is shown for three arrangements of the magnet groups. Also shown are the normalized residuals for each subplot. Vertical black dashed lines indicate the vapor cell face positions. The mean magnetic field along the length of the vapor cell, i.e. between the two black dashed vertical lines, and root-mean-square deviation is stated in each subplot, as well as the magnet group positions.
\label{fig:measured_field}}
\end{figure}

An axial Hall probe was used to measure $B_{z}$ as a function of $z$. Figure~\ref{fig:measured_field} shows three examples of the measured field profile. In each example, the maximum measured magnetic field is distinct due to different positioning of the magnet groups, as discussed in section~\ref{sec:mag_field_comp}. The experimental data is fitted using Eq.~\ref{eq:field}, where we have assumed the magnet dimensions are fixed. We have extracted the positions of the four magnet groups and the remanence field using a Marquardt-Levenberg fit: the errors in the fit parameters are estimated using the square-root of the diagonal elements of the covariance matrix \cite{Hughes2010}. The positions of the magnet groups, the mean magnetic field and root-mean-square deviation within the region of interest, shown via a vertical dashed line, are stated in each subplot. Also shown are the associated normalized residuals for each subplot, which demonstrate excellent agreement between theory and experiment. 

In Fig.~\ref{fig:measured_field}\,a) we produce an axial magnetic field of 1.80(8)~kG and we achieve homogeneity over a length of 25~mm as required. However, we see that we have two local field maxima at approximately $\pm$~40~mm. We attribute these to inappropriate small cylinder positions, and bringing the positions of the small cylinders closer to $z$~=~0 will eradicate these. In spite of this, we have achieved the field required over the region of interest.
As shown in Fig.~\ref{fig:magnet_geometry}\,c), the maximum field we can achieve is $B_\mathrm{{mean}}$~=~2.67(2)~kG due to the hardware constraints, as illustrated in Fig.~\ref{fig:magnet_geometry}\,a). We can achieve any value of field between the maximum and 1.80(8)~kG; examples are shown in Fig.~\ref{fig:measured_field}\,b) and c).

\section{\label{sec:conclusion}Conclusion}

In conclusion, we have presented a permanent magnetic field design that can produce an axial magnetic field of several kilogauss with field homogeneity over a length-scale of tens of millimeters, with negligible contribution from the radial field components. The hardware permits field tuneability ranging from 1.80(8)~kG to 2.67(2)~kG with rms field variation of less than 1\%, and we demonstrate that Marquardt-Levenberg fits to experimental data demonstrates excellent agreement between theory. The design is such that a cylindrical vapor cell of length 25~mm and diameter 25~mm can be placed within the bore of the ring magnets, with a clear aperture of 20~mm for imaging through the vapor cell. The design will be of significance, and benefit, to those interested in atomic spectroscopy in large axial magnetic fields. 
\newline
\section{\label{sec:SuppMat}Supplementary Material}

The supplementary material contains hardware CAD drawings (including drawings for jigs required for assembly and tuning tools), parts list (including a bill-of-materials) for the design discussed here and assembly advice.

\section{\label{sec:ack}Acknowledgments}

We gratefully acknowledge Ifan Hughes and Steven Wrathmall helpful discussions, Clare Higgins for editorial discussions, Stephen Lishman and Malcolm Robertshaw for mechanical design assistance and fabrication, and EPSRC (Grant No. EP/R002061/1) and STFC (Grant No. ST/V005979/1) for funding. 

\section{\label{sec:app}Data Availability}

The data that support the findings of this study are openly available in DRO at https://doi.org/10.15128, reference number r2xs55mc08w \footnote{The data presented in this paper are available from DRO https://doi.org/10.15128/r2xs55mc08w}.

\bibliography{REFERENCES}
\end{document}